\documentstyle[12pt]{article}
\begin{document}

\centerline{\bf How to determine an effective potential for a variable
cosmological term}

\vspace{0.5cm}

\centerline{ A.A. Starobinsky}
\centerline{\it Landau Institute for Theoretical Physics, 117334 Moscow,
Russia}

\vspace{0.5cm}

\noindent
It is shown that if a variable cosmological term in the present Universe
is described by a scalar field with minimal coupling to gravity and with
some phenomenological self-interaction potential $V(\varphi)$, then this
potential can be unambiguously determined from the following observational
data: either from the behaviour
of density perturbations in dustlike matter component as a function
of redshift (given the Hubble constant additionally), or from the
luminosity distance as a function of redshift (given the present
density of dustlike matter in terms of the critical one).

\vspace{0.5cm}
PACS numbers: 98.80.Es, 98.80.Hw

\vspace{0.5cm}

It has been known for many years that the flat Friedmann-Robertson-Walker
(FRW) cosmological model with cold dark matter, a positive cosmological
constant $\Lambda >0$ ($\Omega_0+\Omega_{\Lambda}=1$) and an approximately
flat (or, Harrison-Zeldovich-like, $n_S\approx 1$) spectrum of primordial
scalar (adiabatic) perturbations fits observational data better and has a
larger admissible region of the parameters $(H_0,\Omega_0)$ than any other
cosmological model with both inflationary and non-inflationary initial
conditions (see, e.g., \cite{kof,clas}). Here $H_0$ is the Hubble constant,
$\Omega_0=8\pi G\rho_m/3H_0^2$ includes baryons and (mainly)
non-baryonic dark matter, $\Omega_{\Lambda}\equiv\Lambda/3H_0^2$ and
the light velocity $c=1$. This conclusion was based on the following
arguments: a) relation between $H_0$ and the age of the Universe $t_0$,
b) the fact that observed mass/luminosity ratio never leads to values
more than $\Omega_0 \sim 0.4$ up to supercluster scales, c) comparison
of cosmic microwave background (CMB) temperature anisotropies,
power spectra of density and velocity matter perturbations,
present abundance of galaxy clusters with predictions of cosmological
models with inflationary initial conditions; d) observed values
of $\rho_b/\rho_m$ in rich galaxy clusters confronted with the range
for the present baryon density $\rho_b$ admitted by the theory of
primordial (Big Bang) nucleosynthesis. I don't include gravitational
lensing tests (e.g., a number of lensed quasars) here, since
conclusions based on them are less definite at present; however,
the most recent reconsideration \cite{chiba} has also led to a low
value of $\Omega_0\sim 0.3$.

During last year two new pieces of strong evidence for $\Omega_0<1$
have appeared. The first (historically) of them is based on the evolution
of abundance of rich galaxy clusters with redshift $z$ \cite{bah},
see also the more recent paper \cite{eke} where the value
$\Omega_0\approx 0.5 \pm 0.2$ ($1\sigma$ uncertainty) is presented.
Still, it should be noted that there have been already appeared
some doubts on validity
of the conclusion that $\Omega_0=1$ is really excluded
\cite{blan}. Much better observational data expected in
near futute will help to resolve this dilemma unambiguously. The second,
completely independent argument for $\Omega_0=(0.2 - 0.4)$ follows from
direct observations of supernovae (type Ia) explosions at high redshifts
up to $z\sim 1$ \cite{garn}. On the other hand, no direct evidence
for a negative spatial curvature of the Universe (i.e., for the open
FRW model) has been found. Just the opposite, the latest CMB
constraints (based mainly on the results of the Saskatoon and CAT
experiments) \cite{line}, galaxy abundance at high redshifts
\cite{peac} and the most recent analysis of the SNIa data in terms of
an effective equation of state of a component
adding $\Omega_0$ to unity \cite{garn1} strongly disfavor the
open CDM model without a positive cosmological constant. Of course,
the possibility to have {\em both} a positive cosmological
constant and spatial curvature of any sign is not yet excluded, but,
according to the "Okkam's razor" principle, it would be desirable
not to introduce one more basic novel feature of the Universe
(spatial curvature) without conclusive observational evidence. In any
case, in spite of many theoretical and experimental attempts to
exorcize it, a $\Lambda$-term is back again.

It is clear that the introduction of a cosmological constant requires
new and completely unknown physics in the region of ultra-low energies.
Solutions with a cosmological constant occur in such
fundamental theories as supergravity and M-theory. However,
this cosmological constant is always negative and very large.
As compared to such a basic "vacuum" state, a very small and positive
cosmological constant allowed in the present Universe may be thought
as corresponding to the energy density $\varepsilon_{\Lambda}$ of a highly
excited (though still very symmetric) "background" state. So, it need not
be very "fundamental". But then it is natural to omit the assumption
that it should be exactly constant. In this case the name "a cosmological
term" (or a $\Lambda$-term) is more relevant for it, so I shall use this
one below. The principal difference between two kinds of non-baryonic
dark matter - dustlike CDM and a $\Lambda$-term - is that the latter
one is not gravitationally clustered up to scales $\sim 30~h^{-1}$
or more (otherwise we would return to the problem why $\Omega_0$ observed
from gravitational clustering is not equal to unity). Here $h=H_0/100$
km/c/Mpc.

On the other hand, there exists a well-known strong argument showing that
a $\Lambda$-term cannot change with time as fast as the matter density
$\rho_m$ and the Ricci tensor (i.e., $\propto t^{-2}$) during the
matter dominated stage (for redshifs $z < 4\cdot 10^4~h^2$). Really, if
$\varepsilon_{\Lambda} \propto \rho_m$, so that $\Omega_{\Lambda}=const$,
then matter density perturbations in the CDM+baryon component grow as
$\delta\equiv \left({\delta \rho\over \rho}\right)_m \propto
t^{\alpha} \propto (1+z)^{-3\alpha/2}, ~\alpha= {\sqrt{25-24
\Omega_{\Lambda}}-1 \over 6}$. As a consequence, the total growth of
perturbations $\Delta$ since the time of equality of matter and radiation
energy densities up to the present moment is less than in the absence
of the $\Lambda$-term. If $\Omega_{\Lambda}\ll 1$, then
$\Delta (\Omega_{\Lambda})=\Delta(0)(1-(6.4+2\ln h)\Omega_{\Lambda})$.
Since parameters of viable
cosmological models are so tightly constrained that $\Delta$ may not be
reduced by more than twice approximately, this type of a $\Lambda$-term
cannot account for more than $\sim 0.1$ of the critical energy density
(see \cite{fer} for detailed investigation confirming this conclusion).
This, unfortunately, prevents us from natural explanation of the
present $\Lambda$-term with $\Omega_{\Lambda} = (0.5 - 0.8)$
using "compensation" mechanisms \cite{dolg} or exponential
potentials with sufficiently large exponents \cite{wet}; in other
words, a $\Lambda$-term cannot be produced by an exactly "tracker"
field as was recently proposed in \cite{stein}.

A natural and simple description of a variable $\Lambda$-term is just
that which was so successively used to construct the simplest versions
of the inflationary scenario, namely, a scalar field with some interaction
potential $V(\varphi)$ minimally coupled to the Einstein gravity. Such an
approach, though phenomenological, is nevertheless more consistent and
fundamental than a commonly used attempt to decribe a $\Lambda$-term by a
barotropic ideal fluid with some equation of state. The latter approach
cannot be made internally consistent in case of negative pressure
which is implied by observations \cite{garn1}, in particular, it
generally leads to imaginary values of the sound velocity. On the
contrary, no such problems arise using the scalar field description
(this scalar field is called the $\Lambda$-field below).
Of course, its effective mass $|m_{\varphi}^2|=|d^2V/d\varphi^2|$
should be very small to avoid gravitational clustering of this field in
galaxies, clusters and superclusters. To make a $\Lambda$-term
slowly varying, we assume that $|m_{\varphi}|\sim H_0 \sim 10^{-33}$ eV,
or less (though this condition may be relaxed). Models with a
time-dependent $\Lambda$-term were introduced more
than ten years ago \cite{ozer}, and different potentials $V(\varphi)$
(all inspired by inflationary models) were considered: exponential
\cite{wet,pib,fer,viana}, inverse power-law \cite{rat},
power-law \cite{weiss}, cosine \cite{friem,viana}.

However, it is clear that since we know essentially nothing about physics
at such energies, there exists no preferred theoretical candidate
for $V(\varphi)$. In this case, it is more natural to go from observations
to theory, and to determine an effective phenomenological potential
$V(\varphi)$ from observational data. The two new tests mentioned above
are just the most suitable for this aim. Really, using the cluster
abundance $n(z)$ determined from observations and assuming the Gaussian
statistics of initial perturbations (the latter follows from the paradigm
of one-field inflation, and it is in agreement with other observational
data), it is possible to determine a {\em linear} density perturbation
in the CDM+baryon dustlike component $\delta (z)$ for a fixed comoving scale
$R\sim 8(1+z)^{-1}h^{-1}$ Mpc up to $z\sim 1$, either using the
Press-Schechter approximation, or by direct numerical simulations
of nonlinear gravitational instability in the expanding Universe.
$\delta (z)$ can be also determined from observation of gravitational
clustering (in particular, of the galaxy-galaxy correlation function) as a
function of $z$. On the other hand, observations of SNe at different $z$
yield the luminosity distance $D_L(z)$ through the standard astronomical
expression $m=M+5\log D_L +25$, where $m$ is the observed magnitude,
$M$ is the absolute magnitude and $D_L$ is measured in Mpc.

The aim of the present letter is to show how to determine $V(\phi)$
from either $\delta(z)$ or $D_L(z)$, and to investigate what
additional information is necessary for an unambiguous solution of this
problem in both cases. The idea has been already annouced by the author
in~\cite{star,star1}, now details are given.

The derivation of $V(\varphi)$ consists of two steps. First, the Hubble
parameter $H\equiv {\dot a\over a}=H(z)$ is determined. Here $a(t)$ is
the FRW scale factor, $1+z\equiv a_0/a$, the dot means ${d\over dt}$ and the
index $0$ denotes the present value of a corresponding quantity
(in particular, $H(t_0)=H(z=0)=H_0$). In the case of SNe, the first step
is almost trivial since the textbook expression for $D_L$ reads:
\begin{equation}
D_L(z) = a_0(\eta_0 -\eta)(1+z),~~\eta= \int_0^t{dt\over a(t)}~.
\end{equation}
Therefore,
\begin{equation}
H(z)={da\over a^2d\eta}=-(a_0\eta')^{-1}=
\left[\left({D_L(z)\over 1+z}\right)'\right]^{-1}~.
\label{DLH}
\end{equation}
Here and below, a prime denotes the derivative with respect to $z$. Thus,
$D_L(z)$ defines $H(z)$ uniquely.

More calculations are required to find $H(z)$ from $\delta(z)$. The
system of background equations for the system under consideration is:
\begin{equation}
H^2={8\pi G\over 3}\left(\rho_m + {\dot\varphi^2\over 2}+ V\right)~, ~~~
\rho_m={3\Omega_0H_0^2a_0^3\over 8\pi Ga^3}~,\label{00}
\end{equation}
\begin{equation}
\ddot \varphi +3H\dot\varphi +{dV\over d\varphi}=0~, \label{phieq}
\end{equation}
\begin{equation}
\dot H=-4\pi G(\rho_m + \dot\varphi^2)~. \label{alphaeq}
\end{equation}
Eq. (\ref{alphaeq}) is actually the consequence of the other two equations.

We consider a perturbed FRW background which metric, in the
longitudinal gauge (LG), has the form:
\begin{equation}
ds^2=(1+2\Phi)dt^2-a^2(t)(1+2\Psi)\delta_{lm}dx^ldx^m, ~~l,m=1,2,3~.
\end {equation}
The system of equations for scalar perturbations reads (the spatial
dependence $\exp(ik_lx^l)$, $k_lk^l\equiv k^2$ is assumed):
\begin{equation}
\Phi=\Psi=\dot v~,~~~\dot\delta = -{k^2\over a^2}v +
3(\ddot v+H\dot v+\dot Hv)~,
\label{delta}
\end{equation}
\begin{equation}
\dot \Phi + H\Phi= 4\pi G(\rho_mv+\dot\varphi \delta \varphi)~,
\end{equation}
\begin{equation}
\left(-{k^2\over a^2}+4\pi G\dot\varphi^2\right)\Phi=4\pi G
(\rho_m\delta +\dot\varphi\dot{\delta\varphi}+3H\dot\varphi\delta
\varphi +{dV\over d\varphi}\delta\varphi)~,
\label{Phi}
\end{equation}
\begin{equation}
\ddot{\delta\varphi}+3H\dot{\delta\varphi}+\left({k^2\over a^2}+
{d^2V\over d\varphi^2}\right)\delta\varphi = 4\dot\varphi\dot\Phi -
2{dV\over d\varphi}\Phi~.  \label{dphieq}
\end{equation}
Eq.~(\ref{dphieq}) is the consequence of other ones. Here $v$ and
$\delta\varphi$ are, correspondingly, a velocity potential of a
dustlike matter peculiar velocity and a $\Lambda$-field perturbation
in LG, and $\delta$ is a {\em comoving} fractional matter density
perturbation (in this case, it coincides with
$\left({\delta \rho\over \rho}\right)_m$ in the synchronous gauge).
In fact, all these perturbed quantities are gauge-invariant.

Now let us take a comoving wavelength $\lambda = k/a(t)$ which is
much smaller than the Hubble radius $H^{-1}(t)$ up to redshifts $z\sim 5$.
This corresponds to $\lambda \ll 2000~h^{-1}$ Mpc at present. Then,
from Eq.~(\ref{dphieq}),
\begin{equation}
\delta\varphi \approx {a^2\over k^2} (4\dot\varphi\dot\Phi -
2{dV\over d\varphi}\Phi),~~|\dot\varphi\dot{\delta\varphi}|
\sim |{dV\over d\varphi}\delta\varphi|\sim {a^2H^4\over Gk^2}|\Phi|
\ll \rho_m|\delta|~.
\end{equation}
Therefore, the $\Lambda$-field is practically unclustered at the scale
involved. Now the last of Eqs.~(\ref{delta}) and Eq.~(\ref{Phi})
may be simplified to:
\begin{equation}
\dot \delta = -{k^2\over a^2}v, ~~-{k^2\over a^2}\Phi=4\pi G\rho_m\delta~.
\end{equation}
Combining this with the first of Eqs.~(\ref{delta}), we return to
a well-known equation for $\delta$ in the absence of the $\Lambda$-field:
\begin{equation}
\ddot \delta +2H\dot \delta - 4\pi G\rho_m \delta =0~.
\label{del1}
\end{equation}

It is not possible to solve this equation analytically for an
arbitrary $V(\varphi)$. Remarkably, the inverse dynamical problem,
i.e. the determination of $H(a)$ given $\delta(a)$, is solvable. After
changing the argument in Eq.~(\ref{del1}) from $t$ to $a$ (${d \over dt}=
aH{d\over da}$), we get a first order linear differential equation for
$H^2(a)$:
\begin{equation}
a^2{d\delta\over da}{dH^2\over da}+2\left(a^2{d^2\delta\over da^2}+
3a{d\delta\over da}\right)H^2={3\Omega_0H_0^2a_0^3\delta\over a^3}~.
\end{equation}
The solution is:
\begin{equation}
H^2={3\Omega_0H_0^2a_0^3\over a^6}\left({d\delta\over da}\right)^{-2}
\int_0^a a\delta {d\delta\over da}\, da = 3\Omega_0H_0^2{(1+z)^2\over
\delta'^2}\int_z^{\infty}{\delta |\delta'|\over 1+z}\, dz~.
\label{delH}
\end{equation}
Putting $z=0$ in this expression for $H$, we arrive to the
expression of $\Omega_0$ through $\delta (z)$:
\begin{equation}
\Omega_0=\delta'^2(0)\left(3\int_0^{\infty}{\delta |\delta'|
\over 1+z}\, dz\right)^{-1}~.
\label{omega}
\end{equation}
Of course, observations of gravitational clustering can hardly
provide the function $\delta (z)$ for too large $z$ (say, for $z>5$).
However, $\delta (z)$ in the integrands in Eqs.~(\ref{delH},\ref{omega})
may be well approximated by its $\Omega_0=1$ behaviour (i.e.,
$\delta \propto (1+z)^{-1}$) already for $z>(2-3)$. If massive
neutrinos are present, one should use here the expression with $\alpha$
written above and with $\Omega_{\Lambda}$ substituted by $\Omega_{\nu}/
\Omega_0$ (it is assumed that $\rho_m$ includes massive neutrinos, too).

Finally, using Eq.~(\ref{omega}), Eq.~(\ref{delH}) can be represented
in a more convenient form:
\begin{equation}
{H^2(z)\over H^2(0)}={(1+z)^2\delta'^2(0)\over \delta'^2(z)}
- 3\Omega_0{(1+z)^2\over\delta'^2(z)}\int_0^z{\delta |\delta'|
\over 1+z}\, dz~.
\end{equation}
Thus, $\delta(z)$ uniquely defines the ratio $H(z)/H_0$. Of course,
appearance of derivatives of $\delta(z)$ in these formulas shows that
sufficiently clean data are necessary, but one may expect that such
data will soon appear. Let us remind also that, for $\Lambda\equiv
const$ ($V(\varphi)\equiv const$), we have
\begin{equation}
H^2(z)=H_0^2(1-\Omega_0+\Omega_0(1+z)^3),~~q_0\equiv -1 +
\left({d\ln H\over d\ln (1+z)}\right)_{z=0}={3\over 2}\Omega_0-1~,
\end{equation}
where $q_0$ is the acceleration parameter.

The second step - the derivation of $V(\varphi)$ from $H(a)$ - is very
simple. One has to rewrite Eqs.~(\ref{00},\ref{alphaeq}) in terms of $a$
and take their linear combinations:
\begin{eqnarray}
8\pi GV(\varphi)=aH{dH\over da}+3H^2-{3\over 2}\Omega_0H_0^2
\left({a_0\over a}\right)^3~, \nonumber \\
4\pi Ga^2H^2\left({d\varphi\over da}\right)^2=-aH{dH\over da} -
{3\over 2}\Omega_0H_0^2\left({a_0\over a}
\right)^3~,
\label{V}
\end{eqnarray}
and then exclude $a$ from these equations.

Therefore, the model of a $\Lambda$-term considered in this paper can
account for {\em any} observed forms of $D_L(z)$ and $\delta(z)$ which,
in turn, can be transformed into a corresponding effective potential
$V(\varphi)$ of the $\Lambda$-field. The only condition is that the
functions $H(z)$ obtained by two these independent ways should coincide
within observational errors. $D_L(z)$ uniquely determines $V(\varphi)$, if
$\Omega_0$ is given additionally (the latter is required at the second
step, in Eqs.~(\ref{V})). $\delta(z)$ uniquely determines $V(\varphi)$
up to the multiplier $H_0^2$, the latter has to be given additionally
to fix an overall amplitude. Observational tests which can falsify this
model do exist. In particular, a contribution to large-angle
${\Delta T\over T}$ CMB temperature anisotropy due to the integrated
(or, non-local) Sachs-Wolfe effect presents a possibility
to distinguish the model from more complicated models, e.g., with
non-minimal coupling of the $\Lambda$-field to gravity or to CDM.
However, the latter test is not an easy one, since this contribution
is rather small and partially masked by cosmic variance.

The research was partially supported by the Russian Foundation for
Basic Research, Grant 96-02-17591, and by the Russian Research Project
"Cosmomicrophysics".

\end{document}